\documentclass[aps,jcp,twocolumn,nobibnotes]{revtex4}
\usepackage{graphics,graphicx,amsfonts,amsmath,amsbsy,amssymb,color}
\usepackage{bm}
\usepackage{algorithm}
\usepackage{vector}  
\usepackage{algpseudocode}
\usepackage{paralist}

\def \beq {\begin{eqnarray}}
\def \eeq {\end{eqnarray}}
\def \Schrodinger {{Schr\"{o}dinger }}

\newcommand{\ket}{\ensuremath{\rangle}}

\newcommand{\rff}[1]{{Eq.~\eqref{#1}}}

\def \sym {{\textrm{sym}}}
\def \allow {{\textrm{allow}}}

\usepackage{tikz}
\usetikzlibrary{shapes, arrows, chains, calc}
\usepackage{atbegshi}

\begin{document}
\title{Linear-scaling and parallelizable algorithms for stochastic quantum chemistry}
\author{George~H.~Booth$^{1,2}$}
\email{ghb24@cam.ac.uk}
\author{Simon~D.~Smart$^1$}
\author{Ali~Alavi$^1$}
\affiliation{Chemistry Department, University of Cambridge, Lensfield Road, Cambridge CB2 1EW, UK}
\affiliation{Department of Chemistry, Frick Laboratory, Princeton University, NJ 08544, USA}

\begin{abstract}
For many decades, quantum chemical method development has been dominated by algorithms which involve increasingly complex series 
of tensor contractions over one-electron orbital spaces. Procedures for their derivation and implementation have evolved
to require the minimum amount of logic and rely heavily on computationally efficient library-based matrix algebra and optimized
paging schemes. In this regard, the recent development of exact stochastic quantum chemical algorithms to reduce computational scaling and memory overhead
requires a contrasting algorithmic philosophy, but one which when implemented efficiently can often achieve higher accuracy/cost ratios with small
random errors. Additionally, they can exploit the continuing trend for massive parallelization which hinders the progress of deterministic high-level 
quantum chemical algorithms. In the Quantum Monte Carlo community, stochastic algorithms are ubiquitous but the discrete Fock space of quantum 
chemical methods is often unfamiliar, and the methods introduce new concepts required for algorithmic efficiency. In this paper, we explore these 
concepts and detail an algorithm used for Full Configuration Interaction Quantum Monte Carlo (FCIQMC), which is implemented and available 
in {\tt MOLPRO} and as a standalone code, and is designed for high-level parallelism and linear-scaling with walker number. Many of the 
algorithms are also in use in, or can be transferred to, other stochastic quantum chemical methods and implementations. 
We apply these algorithms to the strongly correlated Chromium dimer, to demonstrate their efficiency and parallelism. 
\end{abstract}
\date{\today}
\maketitle

\section{Introduction}

Post Hartree--Fock methods encompass a range of tools to account for electronic correlations within quantum chemical calculations. 
Such methods are required to progress from a (generally) qualitative description of a system 
at the Hartree--Fock level and approach quantitative agreement with experimental results\cite{Helgie}. 
Recently, three of the most commonly-used methods have been recast in a formalism amenable to stochastic 
evaluation. Each of these has been found to have a number of advantages over their
deterministic counterparts, underlining the potential of these methods.
Stochastic versions of Configuration Interaction (FCIQMC)\cite{BTA2009,BA2010,Nagase2008}, 
Coupled-Cluster (CCMC)\cite{Thom2010}, and M\o ller--Plesset perturbation theory\cite{Thom2007} 
can benefit from reduced computational effort compared to their deterministic 
counterparts, while faithfully reproducing the same results, albeit with small and
systematically controllable random errors, thus maintaining the hallmark of reproducibility in quantum chemical methods. 
The stochastic methods discussed here differ from other recent quantum chemical Monte Carlo schemes based on direct 
energy evaluation\cite{Baer2013,Baer2013_2,Hirata2012,Hirata2013}, as here the wavefunctions are sampled and optimized
in the space of orthogonal Slater determinants.

It may seem counter-intuitive that improvements can be found by removing the large linear algebra routines 
that are so suited to fast computation, but the return comes from the fact that while quantum chemical Hilbert spaces are so 
large, their internal connectivity is relatively small, and there is generally significant sparsity in both the Hamiltonian and the wavefunction\cite{Ruedenberg2001}.
In traditional formulations, deterministic evaluation results in equal computational effort in realizing 
each determinant and transition in the space, regardless of amplitude\cite{Bartlett2007,Knowles1984,Helgie}.
In stochastic analogues low-weighted functions in the space, with few `walkers' residing on them and small
transition probabilites leading to them, consume little computation effort.
Since the low-weighted amplitudes are rarely sampled, each instantaneous snapshot of the walker ensemble represents a
coarse-grained, and highly compressed representation of the wavefunction, with only small parts of the space
instantaneously occupied\cite{BoothDiatomics}. Therefore, the sparsity in the wavefunction can be reflected in the size of the
instantaneous walker distribution, lifting the burden of wavefunction storage which is generally the bottleneck in 
exact diagonalization (FCI) methods\cite{Olsen1988,Knowles1984}.
Nevertheless, time averaging over these instantaneous snapshots within an appropriate dynamic can correctly reproduce the 
wavefunction and energy estimators. 

It should be noted that deterministic schemes to exploit the sparsity in both
the many- and one-electron spaces is a source of much research in wavefunction-based electronic structure theory. 
For example, where the orbitals can be localized, 
cutoffs and local domains are providing a route to take advantage of the generally short-range nature of correlation
and to minimize the redundancy in the space\cite{Chan2012,Neese2013,Werner2011,Maynau2012}. Additionally, tensor factorizations of the wavefunction 
amplitudes aim for an alternative compression of the wavefunction complexity\cite{LowEntanglementWave,Chan2011,Cirac2008,CPS2009,Cirac2009}. 
It is an unresolved and interesting question as to whether the 
stochastic methods could similarly benefit from such localization of the Fock space or tensor network structure imposed on the amplitudes.

Within these stochastic methods, close control over the sampled Hamiltonian gives rise to additional possibilities.
Since Hamiltonian matrix elements are sampled individually, small dynamic modifications and additional criteria on the 
many-body space can give rise to a number of systematically improvable approximations, which can be difficult or
impossible to impose in deterministic methods\cite{CBA2010,CBA2011,BCTA2011,Shepherd2012_2,BoothDiatomics}. 
This approach can again dramatically reduce the computational effort needed to converge to the solution.
In addition to this, stochastic methods can also benefit from improved 
parallelization over distributed memory machines, an important trait on modern computer architecture, and
one which high-level quantum chemical algorithms can particularly struggle with\cite{Evangelisti1993}. 
The efficiency of this parallelism will be explored in this paper. 

In this paper, we focus on the original `initiator' \mbox{$i$-FCIQMC} algorithm. This method has proven successful in providing exact basis-set energies
of systems well outside the limit of what can be achieved within iterative diagonalization schemes in a variety of different systems\cite{BGKA2013,BoothDiatomics,Daday2012,BCTA2011,Shepherd2012_1}.
By reformulating the underlying dynamic of the walker distribution, advances in the scope of the 
method have also been achieved. Complex wavefunctions\cite{BGKA2013}, excited states\cite{Booth2012}, multi-state solutions\cite{Tenno2013,Nagase2010} and finite temperature\cite{BluntArXiv}, as well
as other techniques to reduce the random error or scaling with system size\cite{Clark2012,Umrigar2012} have been developed.
In addition, advances in parent deterministic methods can often
be transferred to their stochastic counterparts, with explicitly correlated versions of the theory and density matrices able to be sampled\cite{BoothF12,OhtsukaF12}.
Many of these methods can be considered as modifications to the underlying walker dynamics of the FCIQMC algorithm from an 
implementational point of view, and so will hopefully also benefit from the careful consideration of this algorithm here.
We shall analyze the performance and implementation of the algorithms which are so critical to the method,
before resolving the electronic effects of the chromium dimer, a molecular system exhibiting 
non-trivial strongly correlated wavefunction structure.

\section{Overview of the FCIQMC algorithm}
\label{sec:Overview}

The key dynamical equations of FCIQMC that are essential to describe 
the algorithm are detailed here, but
a more complete motivation and derivation can be found elsewhere\cite{BTA2009,BCTA2011}. The 
master equation which governs the evolution of the walker population is given by
\begin{equation}
C_{\bvec{i}}^{(n+1)} = \left[ 1 - \tau (H_{\bvec{ii}} - E_0 - S) \right] C_{\bvec{i}}^{(n)} - \tau \sum_{\bvec{j} \ne \bvec{i}} H_{\bvec{ij}} C_{\bvec{j}}^{(n)} .  \label{eqn:master}
\end{equation}
This equation can be derived from a finite-difference formulation of the 
imaginary-time \Schrodinger equation, where the Hamiltonian, $H$, is 
projected into a discrete, orthonormal $N$-electron basis $|D_{\bvec{i}} \rangle$, constructed from a set of $M$ one-electron orbitals. 
The ground-state wavefunction 
in this basis is spanned by the coefficients $C_{\bvec{i}}$ after the master equation is iterated
until convergence at large $n$. The timestep, $\tau$,
represents the discretization of imaginary time, and $S$ is a diagonal
`shift' which is adjusted to maintain a constant number of 
walkers ($L^{1}$-norm). At convergence, this can then be used as an estimate of
the energy.

An alternative single-reference `projected' energy estimator can be 
obtained from 
\begin{equation}
E^{(n)} = \sum_{\bvec{i}} \frac { \langle D_{\bvec{i}} | H | D_{\bvec{0}} \rangle C_{\bvec{i}}^{(n)} }{C_{\bvec{0}}^{(n)}} ,    \label{eqn:ProjE}
\end{equation}
where $|D_{\bvec{0}} \rangle$ defines a reference determinant. This equation is exact
for the correct wavefunction distribution over the functions coupled to
the reference, $\{|D_{\bvec{i}} \rangle\}$, but the random errors associated with this
estimator are sensitive to the weight on this reference function. 
A multireference reference function can be used, and allows for systematic
improvement over the single reference case\cite{Umrigar2012}.

In the standard FCIQMC algorithm, the wavefunction coefficients, $C_{\bvec{i}}$, are now discretized
to an integer representation, where the value of this coefficient on a function is denoted by the number
of signed `walkers' residing there. This is not a unique representation of the wavefunction
coefficients. While some degree of discretization is essential for the compression of the low-weighted amplitudes, a 
multi-scale real/integer representation, where a small region of importance is represented in a
finer, continuous, and deterministic fashion\cite{Umrigar2012}, has been found to provide orders of magnitude saving in the random 
errors in many circumstances. This will not be considered here. 
In this paper, we will focus on 
the algorithm where the wavefunction representation, imaginary time, and off-diagonal dynamic are sampled and fully discretized, 
consistent with the majority of the literature to date.

The dynamics of these walkers now follows a set of steps designed to simulate the evolution of \rff{eqn:master}. First, for each
walker on each determinant, $|D_{\bvec{i}} \rangle$, a symmetry-allowed connected determinant, $|D_{\bvec{j}} \rangle$, is selected at random with a normalized and
calculable probability, $P_{\mathrm{gen}}(\bvec{j}|\bvec{i})$. The {\it spawning} step of the algorithm then proceeds by creating a new signed walker 
on $|D_{\bvec{j}} \rangle$ with
a stochastically realized probability\cite{Mersenne} of
\begin{equation}
    p_{\textrm{spawn}}(\bvec{j}|\bvec{i}) = -sign(C_{\bvec{i}}) \frac{\tau H_{\bvec{ij}}}{P_{\mathrm{gen}}(\bvec{j}|\bvec{i})}   .   \label{eqn:pspawn}
\end{equation}
Where this probability is negative, a particle with a negative sign is created with probability $|p_{\textrm{spawn}}|$. If this probability has magnitude larger than $1$ then the corresponding integer number of particles are created deterministically, and the fractional part stochastically.
After all walkers residing on determinant $|D_{\bvec{i}} \rangle$ have been through a spawning step, a local, diagonal death/cloning step occurs.
This step is applied to all walkers on the determinant at once, and reduces the population on the local function with another stochastically
realized probability of
\begin{equation}
p_{\textrm{death}}(\bvec{i}) = \tau C_{\bvec{i}} (H_{\bvec{ii}} - E_0 - S)    .   \label{eqn:pdeath}
\end{equation}
Where $p_{\textrm{death}} < 0$ anti-particles are spawned which grow the population. As generally $H_{\bvec{ii}} \ge E_0$ anti-particles are only spawned for large, positive values of $S$.
It should be noted that walkers spawned, cloned or killed within an iteration do not contribute to the subsequent steps of the same iteration.

\begin{figure*}
	\begin{center}

        \includegraphics[scale=0.9]{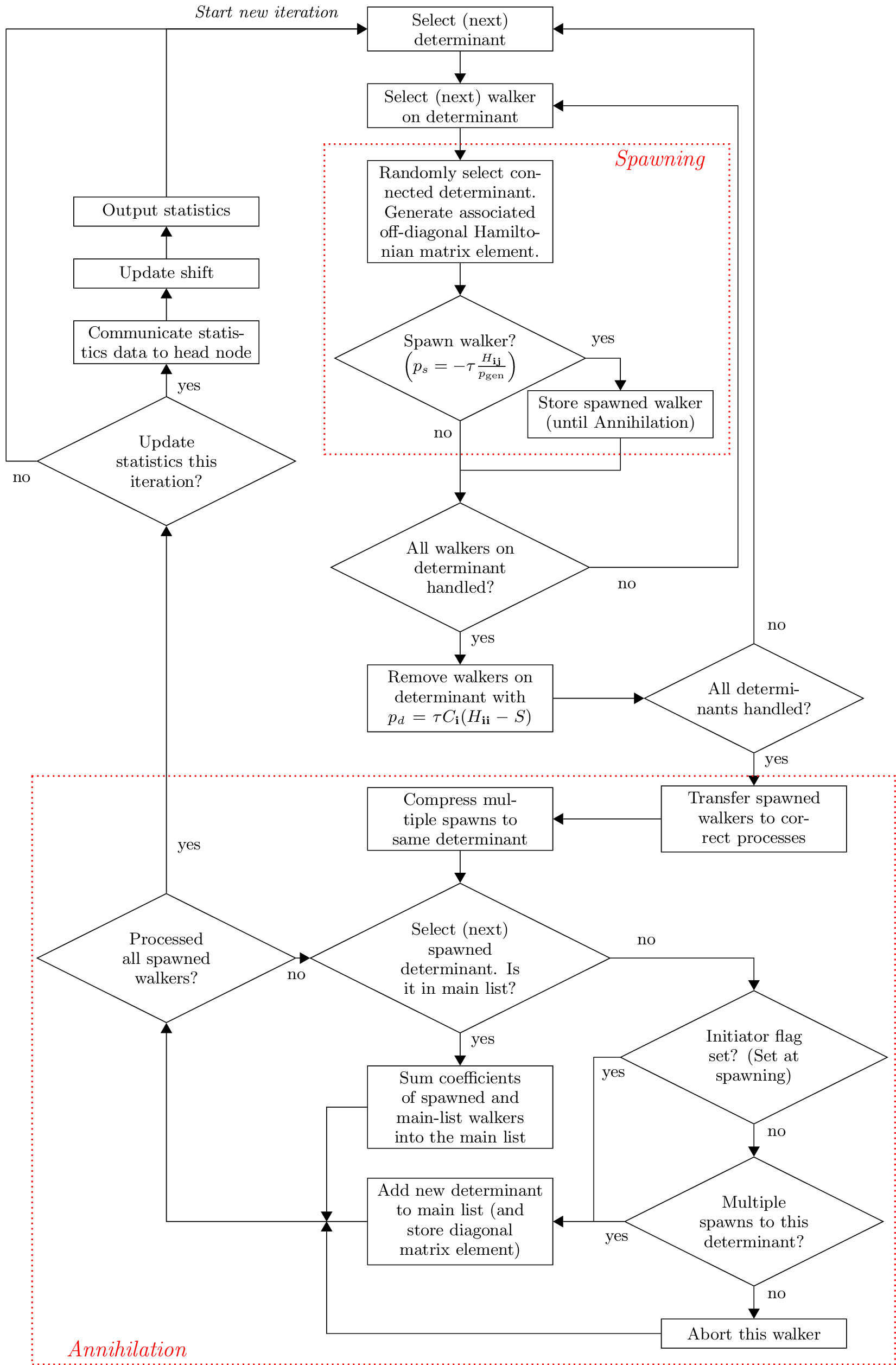}
		\caption{Overview of the FCIQMC algorithm, showing spawning, death and annihilation steps of the main iteration loop.}
		\label{fig:FCIQMC-algo}
	\end{center}
\end{figure*}

Taken together, these two steps simulate the dynamics of \rff{eqn:master}. However, for general Fermionic wavefunctions, it is impossible to 
find a representation of the wavefunction such that all $C_{\bvec{i}}$ amplitudes are of the same sign\cite{Troyer2005}. This results in a propagation of
both positive and negative walkers, and a manifestation of the Fermion sign problem within this algorithm. Although this has been shown
in general to be less severe than the analogous problem within real-space QMC approaches\cite{Foulkes2013}, the FCIQMC algorithm can exactly overcome this
exponential reduction in signal to noise ratio via local annihilation events between oppositely signed walkers on the same 
determinant\cite{BTA2009,Spencer2012}. An 
important consequence for this is that the walkers on each determinant at the beginning of an iteration are all of the same sign. 
It is also at this annihilation stage that additional approximations such as the successful `initiator' adaptation of the method can be applied.
The annihilation algorithm will be detailed in section~\ref{sec:Annihilation}. A flow diagram detailing the main loop and logic structure of the
overall algorithm is given in Fig.~\ref{fig:FCIQMC-algo}.

\tikzstyle{base} = [draw, on chain, on grid, align=center, minimum height=4ex]
\tikzstyle{block} = [base, rectangle, text width=10em]
\tikzstyle{test} = [base, diamond, aspect=2, text width=7em]
\tikzstyle{normal-arrow} = [->, draw, black]
\tikzstyle{reverse-arrow} = [<-, draw, black]
\tikzstyle{normal-plain} = [-, draw, black]
\tikzstyle{coord} = [coordinate, on chain, on grid]
\tikzstyle{narrow-coord} = [coord, node distance = 6mm and 25mm]
\tikzstyle{norm-coord} = [coord, node distance = 6mm and 30mm]
\tikzstyle{wide-coord} = [coord, node distance = 6mm and 52mm]
\tikzstyle{wider-coord} = [coord, node distance = 6mm and 55mm]

The innermost loops of the FCIQMC algorithm (the spawning steps) involve random generation of symmetry-allowed connected 
determinants, and the calculation of the Hamiltonian matrix elements which connect the two. The generation of excitations is considered
in section~\ref{sec:Excitgen}, while the generation of matrix elements follows standard Slater--Condon rules (for a determinant basis)\cite{Helgie}.
A substantial cost in the algorithm, when using large basis sets, is memory latency in the one- and two--electron integral lookup between arbitrary orbitals
required for these matrix elements. 
Since integrals are required in a random order, a pre-fetching
algorithm to obtain multiple integrals at once is difficult to implement, while the $\mathcal{O}[M^4]$ number of the integrals is likely
to provide the memory bottleneck in larger studies\footnote{A double triangular mapping reduces this number by a factor of eight to 
account for the permutational symmetry of the integrals, or four in the case of complex orbitals.}. 
This is somewhat ameliorated by a shared memory implementation 
(via either {\tt POSIX} or System V shared memory\cite{UNIX1997,Robbins2003}), but for large number of
orbitals ($\gtrsim 250$), either a density-fitted/Cholesky decomposed integral engine\cite{Ahlrichs2009,Martinez2012} (to reduce 
integral storage to $\mathcal{O}[M^3]$ or lower), on-the-fly 
calculation, distribution and communication of integrals between computational nodes or other compressed representation 
will likely be necessary. However, this will serve to increase the integral lookup cost and is not considered further here. In all applications
to date, the integrals have either been stored in memory on all computational nodes, or calculated on-the-fly.

The other algorithmically and computationally non-trivial step required, which will be unfamiliar to standard deterministic quantum chemistry
packages and is key to the performance of the method, is an efficient and scalable annihilation algorithm between walkers on the same function
with opposite sign. Much of the implementation of the algorithm is with this in mind, since naive approaches can be very costly. This 
is also the only step which involves communication between MPI threads other than occasional aggregation of statistics, and is key to 
the parallel performance of the algorithm. It is important, therefore, to perform the MPI part of this 
step in a single, collective operation to reduce communication latency. 
The implementation of this step will be considered in section~\ref{sec:Annihilation}.

Unless such large numbers of parallel processes are used that communication becomes the bottleneck, we have found that excitation
generation is the most costly part of the algorithm. This is dependent on the number of irreducible representations in the symmetry
group being used within the system ($N_{\textrm{sym}}$), as the cost increases with additional symmetry elements. However, 
we have found it always worthwhile to make use of symmetry where possible, as this generally has a quadratic saving, since
both the size and the internal connectivity of the space are reduced by a factor of the number of symmetry elements, resulting in an increase
in the $p(\bvec{j}|\bvec{i})$ values obtained in the spawning step and a corresponding increase in the timestep that can be used. Abelian subgroups
of D$_{\infty {\mathrm{h}}}$ are available, as are full L$_{\mathrm{z}}$ symmetries, translational point groups, total spin eigenfunctions and 
time-reversal symmetries\cite{Larsen2000,BCTA2011,Nagasehphf}.

Other factors such
as the number of walkers, orbitals, electrons and computational cores run on, as well as effects such as the sparsity of the wavefunction will
all influence the efficiency of the algorithm and may change the limiting step. The algorithm will generally become 
increasingly parallelizable with increasing walker number, as needed to converge the energy for larger systems. 
Because of this, we will focus on the computational efficiency in the large walker limit of the algorithm, by outlining a 
scheme for linear scaling with respect to the number of walkers, $N_w$, where availability of
memory is not a bottleneck. Derivatives of this scheme can be employed in other computational regimes, where speed can be sacrificed for memory saving, 
although these schemes will not be discussed here as memory availability is generally not the bottleneck of this algorithm on modern computational resources (as
opposed to deterministic FCI).

\section{Representation of walkers}

The most simple $N$-electron space of the FCIQMC dynamic is the complete set of Slater determinants, however there are other function spaces which can 
span the same Hilbert space, while being more compact. These include fixed combinations of determinants which obey spin-reversal symmetry (see 
section~\ref{sec:hphf}), momentum-reversal symmetry, or
total spin eigenfunctions (configuration state functions\cite{Helgie}). Working in these other spaces involves a 
trade-off between the complexity of the excitation
generation algorithms and matrix element evaluation, and the advantages provided by reductions in the total size and connectivity of the 
spaces, or when aiming separately for ground and excited states which were previously of the same symmetry.
When `determinants' are mentioned here, it should be implicit that any of these function
spaces can be used in its place. Indeed, more complicated function spaces, including non-orthogonal or geminal spaces may also provide
interesting research directions in the future. 

It is important for the FCIQMC algorithm that all of these functions can be uniquely and compactly represented by bit strings indicating the occupations
of the $M$ orbitals (or $2M$ spin-orbitals) in the space. For determinants, this is straightforward and has been in use in previous `string'-based
schemes\cite{Knowles1984,Lin1990}. For spin-coupled spaces, one unique determinant from the coupled pair is chosen to denote 
the function, so that the representation is always the same. Many operations on determinants are then reduced to bit operations. For instance,
finding the number of orbitals differing between two determinants can be computed via an `exclusive or' operation (XOR), followed by counting 
the set bits of the result\cite{AndersonWeb}. Additionally, checking whether two functions are the same is equivalent to testing the equality of the 
bit-representations.

In our implementation, the bit-string is formulated from an integer representation, and multiple integers are used when the number of spin-orbitals exceeds
the number of bits in the integer type. In addition, another single integer is used to store the signed 
number of walkers occupying the determinant, and any single bit `flags' which may be required, such as whether a walker has 
been spawned from a function which is deemed an `initiator' (which confers 
special properties to the walker regarding its survival if spawned to an unoccupied determinant\cite{CBA2011,BCTA2011}). 
To isolate the number of walkers on a determinant, their sign, or the associated flag, separate masking integers are used to
isolate the component bits of this integer when desired via AND operations. This compression of multiple data into a single integer is done primarily to
minimize the amount of data for communication purposes, rather than to save on memory usage. For complex wavefunctions for use with complex
irreducible representations such as translational group symmetry and crystal momentum for solid-state systems, or for other complex wavefunctions
found in e.g. systems with spin-orbit coupling, two integers are used for the walker population to denote separately the real and imaginary parts of the
wavefunction coefficient.

This representation of the walkers as a set of integers uniquely denoting the determinant, signed number of walkers, and any flags conferring special attributes
is the standard representation for all walkers in both the main list, and the list for newly-spawned walkers each iteration. The total number of 64-bit 
integers required to store an occupied determinant in the main list therefore scales with number of orbitals as $\lceil \frac{2M}{64} \rceil + 1$. By way
of illustration, in a system with 128 spin-orbitals, 100Mb of memory will therefore store over four million occupied determinants, and assuming an
optimal load-balancing (see section~\ref{sec:Annihilation}), this number of distinct occupied determinants can be stored on each computational process on
distributed memory architecture. It should be stressed that the storage of the determinants is only ever performed over the instantaneously occupied
determinants, and as such, no memory requirements which explicitly scale with the size of the full Hilbert space are ever required. 

Finally, it can be useful to store the diagonal Hamiltonian matrix elements for the occupied determinants in the main list, along with the 
standard (integer) determinant representation. This is not essential, but saves on regeneration of these matrix elements, which involves
an $\mathcal{O}[N^2]$ operation, at each death step if a determinant remains occupied over multiple iterations.
Crucially, this data does not need to be communicated in the annihilation step, and therefore involves only a memory 
cost rather than increasing the quantity of data to communicate.

\subsection{Encoding and Decoding of Determinant Representations}

Encoding of a newly occupied determinant bit-string representation from scratch is rarely required. 
Since determinants are generated through excitations of already occupied determinants, the bit representation of the new determinants 
may be calculated from the old in an $\mathcal{O}[1]$ step by clearing the bits representing the source orbitals 
and setting those representing the targets. However, in the excitation generation step, and for certain operations such as Hamiltonian
matrix element calculation for single excitations and hashing (see section~\ref{sec:Annihilation}), an alternative determinant representation is preferable.
In this `electron occupation' representation, an ordered set of $N$ integers are used to specify the spin-orbitals occupied by 
each of the electrons in the function. This representation is generated for each occupied determinant from its bit representation 
via a 1-to-1 `decoding' function when it is considered in each iteration.

Decoding can be performed naively by looping over all the bits in the bit representation, and appending each set bit to a list. 
However this is unnecessarily costly.
In a similar way to one method of counting the number of bits set in an integer\cite{AndersonWeb}, this may be approached through 
subdividing the bit representation into individual bytes, and creating a lookup table for the available 256 possibilities\footnote{We would like to thank James Spencer for suggesting this approach, private communication.}. 
Each possible byte has an entry containing 
\begin{inparaenum}[\itshape i\upshape)]
	\item the number of orbitals contained in this byte, and
	\item a list of these orbitals (with the first bit in the byte being orbital zero).
\end{inparaenum}
Looping over all non-zero bytes, until the correct number of orbitals are found, is substantially more efficient than looping over each of the bits.

\section{Hashing and Annihilation}
\label{sec:Annihilation}

The annihilation of walkers of different signs on the same determinant is of crucial
importance to the emergence of the sign structure of the wavefunction\cite{BTA2009,Spencer2012,Foulkes2013}. Since
only the instantaneously occupied determinants are stored, rather than a histogram of the whole
determinant space, this annihilation step has to be performed explicitly.
A dual hashing procedure is a key feature of this FCIQMC algorithm, on which rests the load-balancing and parallelism of the algorithm 
over the available computational processes, as well as the linear scaling with respect to walker number. 

A hash function is a many-to-one mapping from a data set to a (in this case) single 
integer within a predetermined range. Generally, it is simple to map from the data to the hash value, but very difficult to perform the reverse and
is hence used for encryption, although this is not a feature of the hash function which will be exploited here. Instead,
this algorithm is dependent on a uniform distribution of hash values across the full range desired, with emphasis on
low-order bits changing rapidly and fast evaluation.
To this end, a simple Merkle-Damg\aa rd type hash outlined in Algorithm~1 has proved useful in mapping a determinant in electron occupation 
representation to a single integer\cite{Merkle89,Damgard89}. It is possible to directly hash the bit-string determinant representation, although this compressed form leaves
less data on which to perform the hash, producing less uniformly distributed results. Therefore it was found to be preferable to use the electron occupation representation in cases of small orbital basis sizes.

Common orbital orderings will order the orbital indices by energy, or by symmetry. Either way, there is likely to be significant common structure in the
representation of the dominant determinants in the wavefunction, including structure from the global M$_{\mathrm{s}}$, and the generally occupied core
orbital configurations. To somewhat mitigate these effects in the resulting hash, an additional simple 1-to-1 mapping is made from the 
spin-orbital indices to another set of
random integers, whose range can be much larger than the original number of spin-orbitals. This additional random lookup table increases the 
entropy in the data set and results in more uniform hash values.

\begin{algorithm}
\begin{algorithmic}
\State $hash = 0$
\For{$i = 1 \to nElec$}
\State $hash \gets p \times hash + \mathrm{map(det(}i\mathrm{))} \times i$
\EndFor
\State $hash = \mathrm{abs(mod(}hash,range))$
\end{algorithmic}
\label{code:Hash}
\caption{Simple FNV hash algorithm to return a hash value ($hash$) in the range $0 \rightarrow range-1$, from an $nElec$-electron 
determinant in electron occupation representation ($det(1:nElec)$). $p$ represents a large prime number (we currently use 1099511628211), 
while $map$ is a simple 1-to-1 mapping function
from a specified spin orbital to a unique integer from a large range, designed to increase the entropy from the available information. Integer overflow
is likely and desired.}
\end{algorithm}

\subsection{Hashing for parallel performance}

On each parallel process, a main list stores the list of occupied determinants on that particular process, with their signs and flags,
disjoint to all determinants stored on other processes.
The process a determinant is assigned to is determined from the hash value across the range
of the number of processes.
The occupied determinants (not individual walkers) are therefore distributed across the different MPI processes (which generally 
correspond to the computational processing cores). Each determinant is found on the same process at all stages of the calculation.
It is crucial
to be able to deterministically compute which process this is, such that newly spawned walkers can be communicated to the correct process
for the annihilation step. 

\begin{figure}  
\begin{center}
\includegraphics[scale=0.4]{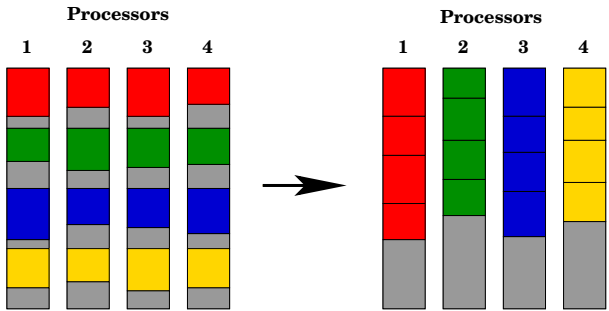}
\end{center}
\caption{Schematic to represent the movement of walkers for the inter-process communication required for the annihilation step. The 
non-contiguously filled array containing the 
newly spawned walkers is shown for each process on the left before the communication, while the right is the set of walkers
on the correct process after a single, collective communication step per iteration. The load imbalance in the number of spawned
walkers per process is of minor concern compared to the balancing of the number of determinants in the main lists due to the fact that
$N_s \ll N_w$, and that the annihilation routine is only of cost $\mathcal{O}[N_s]$. 
}
\label{fig:Annihil}
\end{figure}

During the iteration, each determinant in the main walker array is considered (see section~\ref{sec:WalkerListHash}), and 
attempts spawning and death steps.
Each time a new walker is spawned, its hash value is computed between the range of available
computational processes, giving its target process, and the bit-representation stored in a non-contiguous `spawned' array, separate to the main walker array, on the process on which it was created.
This additional spawned array of walkers only needs to store the maximum number of successful spawning events on each process per iteration ($N_s$), 
and so can be much smaller than the main array, by a factor of 10-1000 ($N_s \ll N_w$). The position in the array is determined by the hash, and therefore 
non-contiguously orders the newly spawned walkers by the
process on which the determinant should be located after communication, as shown on the left figure in Fig.~\ref{fig:Annihil}. 
Local death/cloning events are however updated directly in the main array.

At the end of the main iteration loop, the newly spawned walkers are sent to their designated processes via a single, synchronized, collective
operation (within the MPI library, an {\tt MPI\_AlltoAllv} operation). As illustrated in Fig.~\ref{fig:Annihil}, due to the way that the
newly spawned walkers were ordered, this takes the form of a non-contiguous matrix transpose. After this, each process' list of 
newly spawned walkers only includes walkers assigned to that process. As a consequence, all walkers on any given determinant
will now reside on the same process.
This allows annihilation to occur fully without any further communication.
After the movement of walkers, the small, now contiguous newly spawned walker list on each process is ordered and compressed so that there are no 
multiply specified determinants in the list\footnote{This step is not essential, and can be omitted with care. Assuming a sensible sorting routine, the
cost is $\mathcal{O}[N_s \log N_s]$, which strictly violates the linear scaling with respect to $N_w$. However, since $N_s \ll N_w$, this cost
can be considered small, and it is practically favorable to perform it.}. 
This combines newly spawned walkers of the same sign into one entry in the list, and 
locally annihilates walkers of different sign between those spawned in the current iteration.
The mechanism of further annihilation with entries from the main list on the process will be described in section~\ref{sec:WalkerListHash}.

\begin{figure}  
\begin{center}
\includegraphics[scale=0.475]{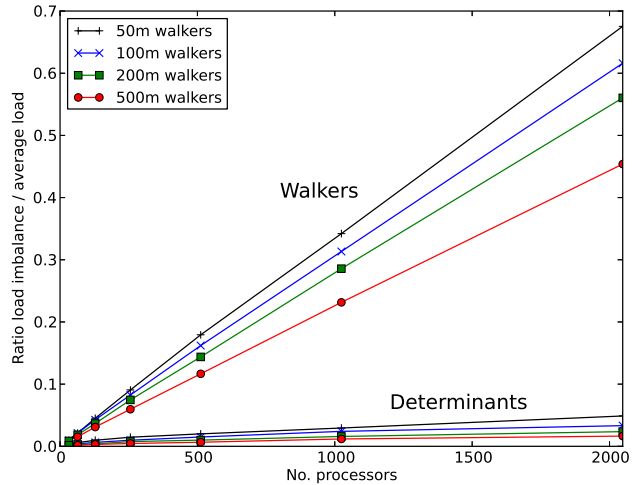}
\end{center}
\caption{The observed difference in the number of walkers and occupied determinants between the
most and least occupied processes relative to the average number per process. It is notable that
the distribution of determinants is substantially better balanced than the distribution of walkers.
In this case, the absolute difference in walker count between the most and least occupied process
is roughly equal to the number of walkers on the reference (Hartree--Fock) determinant, and this
comes to dominate as the number of processes is increased, reducing the average occupancy of each
process - an effect which dominates the parallel scaling.}
\label{fig:LoadBalance}
\end{figure}

The load-balancing of the algorithm across processes is therefore tied to the uniformity of the hashing function. A beneficial property 
of the algorithm is the fact that 
the larger the system ($N$), the more entropy that is available to the hashing function, and therefore the more uniform the hashing
should become. In practice, the load balancing of the hash values is rather uniform and is not the bottleneck of the parallelism. 
However, this is not the only consideration, as this algorithm balances
the number of occupied determinants across the available processes rather than the number of walkers, and the equal balancing of computational
effort will depend on both of these quantities, as can be seen in Fig.~\ref{fig:LoadBalance}. 

It is not clear which it would be 
better to balance, since some operations such as the death
step, diagonal matrix element evaluation and annihilation take place on the level of occupied determinants, while others such as the spawning attempts and
the majority of the excitation generation take place on an individual walker level. 
The optimal load balancing would include some of both characters, but this
algorithm is limited to only consider balancing of occupied determinants only across computational processes. Therefore, if a walker distribution
is heavily skewed towards, say, the Hartree--Fock determinant, then this will result in a load-imbalance on the level of walker distribution across
computational processes.

\begin{figure}  
\begin{center}
\includegraphics[scale=0.475]{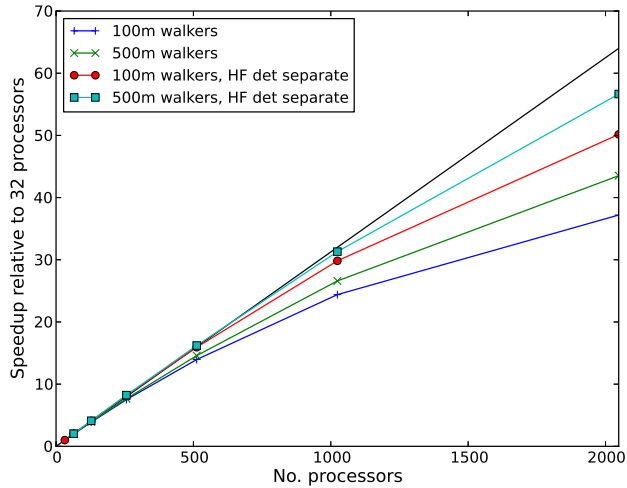}
\end{center}
\caption{The observed speed-up as the number of computational processes is increased from 32 to 2048.
We note that the scaling improves as the number of walkers in the system is increased. Empirically
we observe that the computational speed scales fairly efficiently so long as at least 1 million walkers
are present on each computational process. In systems where the walker distribution is skewed to a
single occupied determinant, such as the Hartree--Fock determinant, it is advantageous to the
parallelism of the algorithm to assign a dedicated process to the storage of this determinant, as shown.}
\label{fig:Scaling}
\end{figure}

Despite this limitation, an upside of this is that for a given number of processing threads, increasing the number of walkers will lead to an increased
parallel performance as a larger fraction of the space is simultaneously occupied and the load is more evenly balanced as can be seen in Fig.~\ref{fig:Scaling}. 
This means that as systems become larger and more walkers are required for convergence to the correct energy, the parallelism of the algorithm increases, allowing it to be run more
efficiently on larger computational resources. However, it is possible within the algorithm to isolate subsets of determinants (which need to be readily identifiable)
which can be separately considered on a dedicated process if they are particularly computationally intensive, while the remainder of 
the determinants are placed based on their hash over a now reduced range of available processes. The simplest of these schemes involves placing just the largest weighted determinant
on a dedicated computational processes, while distributing the others accross the remaining processes. The effects of this can be seen in Fig.~\ref{fig:Scaling}, where the Hartree--Fock
determinant has a seperate process, and no dynamic load-balancing is performed. As can be seen, this greatly improves the parallelism of the algorithm. Schemes to improve this
further will be considered in the future, where the true computational load is balanced as evenly as possible.

\subsection{Hashing for main walker list}
\label{sec:WalkerListHash}

The second hashing is not for the purposes of parallelism and involves the explicit construction of a hash table for the storage of the main walker list
on each process. The idea is for the hash value of a determinant
to return a position in a hash table, which will in turn provide the array index for the determinant in the main walker list on the local process. The
reason for the hash table construction is that the time taken to search for a determinant in the main list, insert a 
new determinant or delete an old determinant is $\mathcal{O}[1]$.
Two integer arrays in memory are required in addition to the main walker list. 
The first is the hash table, which can be of variable length, but should optimally scale with the size
of the main determinant array to minimize hash collisions. The other is a circular array indexing the vacant positions in the main determinant list into
which new determinants can be placed. This latter list has a maximum size given by the length of the main determinant list, but can be shorter
\footnote{In a previous incarnation of the algorithm, the main walker array was kept ordered and contiguous, and searching 
for a determinant in the main array was then performed via a binary search. This algorithm can still be used, as it will require less memory (no need to
store the intermediary hash table or circular list). However, it is computationally more expensive for large numbers of walkers as the annihilation step scales formally 
as $\mathcal{O}[N_s \log N_d]$ (rather than $\mathcal{O}[N_s]$) where $N_d$ is the number of occupied determinants. It also requires more 
costly insertion/removal steps for changing determinant lists.}. 

In the main loop, the occupied determinant list is sequentially accessed. Each time an empty entry in the list is found, the position index
is saved in the circular list. Spawning and death steps are performed as normal for entries corresponding to occupied
determinants in the list. If all walkers on a determinant are killed during the death step, then its walker count is set to zero, and its index is added
to the circular list and removed from the hash table. Once all walkers on the process have been considered, there is no need to
continue iterating through the main walker array, and all further (vacant) entries in the main determinant list can be
considered available for possible walker insertion.
In this way, the first entries filled from determinant insertion are the earliest vacant entries, meaning that the main
determinant list remains as 
contiguous as possible, minimizing the number of empty array entries which are searched through.

During annihilation, when the contiguous newly spawned walker list is compared to the existing determinants in the main walker list, the hash
value of each determinant is computed, and the index of the determinant in the array looked up from the hash table. If there are hash collisions
(two determinants with the same hash), then 
multiple entries in the main list may need to be considered before the correct determinant is found. If the determinant is found, then annihilation or
addition to the existing entries can be achieved simply by modification of the number of walkers in the corresponding entry in the main list.
If the resultant number of walkers is zero, the entry should be removed from the hash table, and the associated 
index added to the circular list of 
free slots available for determinant insertion, in the same way as for the death of all walkers on a determinant. 
Alternatively, if no corresponding determinant is found, the newly spawned walker can be
transferred to the main list in the next free position indicated in the circular list.
In this way, the entire annihilation step
can be efficiently performed in only $\mathcal{O}[N_s]$ computational cost.

The reason for the intermediary hash table step to provide an index is due to the added simplicity when dealing with hash collisions.
This is not essential and could be removed to provide memory savings. Assuming uniformity of the hash values, the probability of a hash
collision between two determinants on the same process is given by
\begin{equation}
p_{\textrm{collision}} = 1 - \frac{N_h!}{(N_h - N_d)! N_h^{N_d}}    ,
\end{equation}
where $N_h$ is the size of the hash table, and $N_d$ denotes the number of occupied determinants on the process. 
It can be seen in this way that increasing the size of the hash table is advantageous for minimizing the risk of hash collisions.
In the case of a hash collision,
the additional distinct determinant index is included in the same entry in the hash table, resulting in multiple entries potentially being searched
at the annihilation stage. Further hash collisions are possible, but become scarcer. It is also worth mentioning that different random mapping arrays
are used in the hashing functions for the distribution over processes and over the main determinant array (as well as different ranges), in order
to avoid any unwanted correlations being introduced between the two hashes.

%
It is during this annihilation stage that the initiator criterion on newly spawned walkers can be imposed with virtually no overhead. 
For any walker in the spawned list, if no corresponding determinant is found in the main list when searched for,
the walker's flag is tested to determine if it was spawned from a determinant deemed to be an `initiator'.
If so, the insertion into the main determinant list proceeds as normal. Otherwise, the spawned walker is discarded from the simulation.
Similarly, it is also
at this stage where the diagonal Hamiltonian matrix elements are calculated (if they are being explicitly stored), if a new determinant is being
occupied. 

Another advantage of this algorithm is that for the lifetime of a particular determinant, it can always be found in the same position in the main
walker array. If the weight on particular determinants, such as the Hartree--Fock or $T1$ amplitudes in CCMC, need to be accessed frequently 
then it is not necessary to search the array to locate them, but rather their indices can be stored over their lifetime.
This also eliminates memory `churn' incurred from moving walkers around the main list to maintain contiguousness.

\section{Random excitation generation}
\label{sec:Excitgen}

A non-trivial part of these stochastic quantum chemical methods is concerned with the random selection of
connected determinants, within the symmetry constraints imposed upon the space.
The primary difficulty associated with the construction of the excitation algorithm is that the generation probability for each excitation
must be computable, with the sum of all possible outcomes for each source
determinant correctly normalised, and all possible routes for generating 
each resultant determinant included in any calculated probability.
This is so that the overall spawning probability (Eq.~\ref{eqn:pspawn} multiplied by the probability of selecting the excitation)
for any transition is independent of the specifics of the random excitation process, and purely a linear function of the connecting
Hamiltonian matrix element. 
It is this normalization criterion which affects the efficiency of the excitation generation, since explicit normalization by full 
or even partial enumeration of all excitations is too costly as system sizes increase. 

The generation of normalized probabilities does not mean that excitations must be generated in a uniform manner. 
In addition, in this algorithm, it is possible to return a ‘null’ or aborted determinant from the excitation routines. If a null excitation is generated,
the Hamiltonian matrix element to this excitation is considered zero, and no spawning can occur.
Additional restrictions can be imposed on the allowed space to search, by returning null determinants when determinants outside the allowed
space are generated
\footnote{Although truncated/active space calculations are currently performed by returning null excitations outside the permitted space, it would also 
be possible to adapt the algorithm to only compute excitations up to a specific excitation level directly by considering only allowed classes of excitations.
This would lead to a more efficient sampling of the excitation space and higher acceptance ratios for the same timestep.}.
This trivially allows for truncated or active space CI calculations, or truncations according to other criteria such as seniority 
number\cite{Bytautas2011}.

\begin{figure*}
	\begin{center}
		\includegraphics{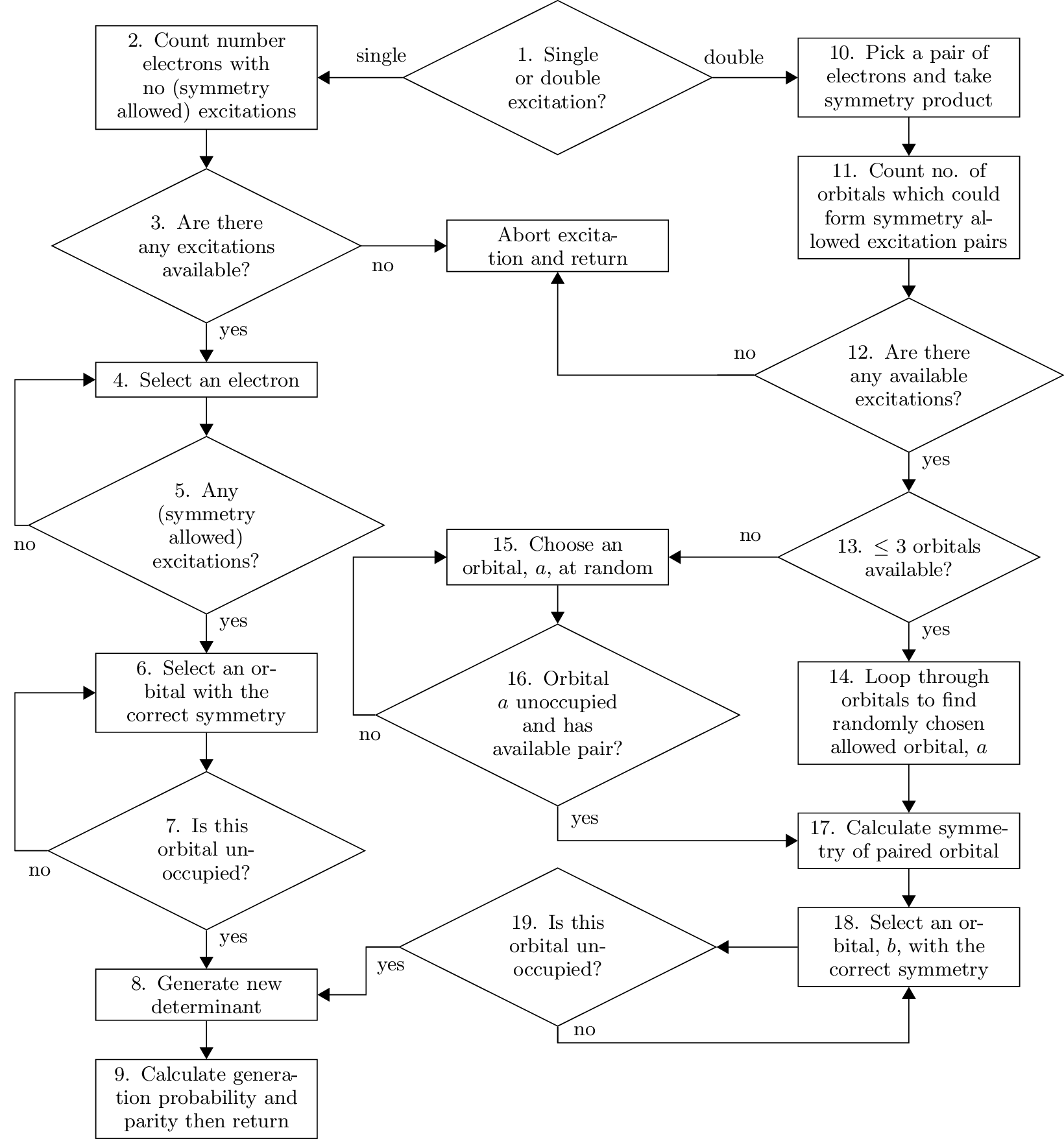}
		\caption{Random excitation generation overview}
		\label{fig:excit-gen}

	\end{center}
\end{figure*}

A flow diagram, giving the main steps in our excitation generation algorithm is shown in Fig.~\ref{fig:excit-gen}. Overall, the aim is to return
a singly or doubly excited determinant of an initial determinant, $|D_{\textrm{src}} \rangle$. A singly excited determinant involves finding an occupied
orbital $i \in |D_{\textrm{src}} \rangle$ and unoccupied orbital $a \notin |D_{\textrm{src}} \rangle$, while a double excitation requires an occupied
orbital pair $\{i,j\} \in |D_{\textrm{src}} \rangle; i \neq j$ and unoccupied pair $\{a,b\} \notin |D_{\textrm{src}} \rangle; a \neq b$ to be selected. 
An initial step creates two integer lists, detailing the number of occupied and unoccupied spin-orbitals of each symmetry in the
determinant $|D_{\textrm{src}} \rangle$. These lists can be trivially computed in $\mathcal{O}[N]$ time, and since they are the same for all excitations from
$|D_{\textrm{src}} \rangle$, the operation is performed once, and then saved between multiple excitations from walkers on the same determinant.
Symmetry in this sense refers to all one-electron symmetry labels on the orbitals, including potentially 
a spin label, point group irreducible representation, $m_l$ quantum number (if conserving $L_z$ symmetry\cite{BCTA2011}), and $k$-point labels 
in systems with translational invariance\cite{BGKA2013,Shepherd2012_2,Shepherd2012_1}.

Upon attempting to generate an excitation, the first choice is which type of excitation (single or double) to generate,
as shown in Fig.~\ref{fig:excit-gen} as (1).
This choice does not need to reflect the exact ratio in the number of each type, though it always helps to be as close
as possible to encourage uniformity in the generation probabilities and improve sampling. The choice however does need to maintain the
normalization, requiring
\begin{equation}
    P_{\textrm{double}} + P_{\textrm{single}} = 1 .
\end{equation}
By assuming the ratio of double to single excitations should be roughly constant across the determinant space, $P_{\textrm{double}}$ and $P_{\textrm{single}}$ are
fixed throughout the run, after choosing them initially by explicitly calculating the ratio of excitation types from a Hartree--Fock or other reference determinant.
The quality of this assumption is dependent on the uniformity in the number of irreducible representations. 
The first stochastically realized choice in the excitation generation is therefore based on the probability $P_{\textrm{double}}$.
Note that in some model Hamiltonians (e.g. the Hubbard model or uniform electron gas in direct or reciprocal space), only double or single excitations
are allowed, and this initial selection can be avoided.
    
\subsection{Single Excitations}

In small systems, or systems with a high degree of symmetry, it is possible to choose an electron to excite, which has
\emph{no} symmetry allowed excitations from it. For a single excitation to be symmetry allowed, it requires
\begin{equation}
    \Gamma_i \otimes \Gamma_a \ni A_1   ,
\end{equation}
where $\Gamma_i$ and $\Gamma_a$ represent the irreducible representations of orbital $i$ and $a$, and $A_1$ is the 
totally symmetric representation, which generalizes to spin and other symmetries.
If no $a$ orbitals match this criterion for a randomly chosen $i$, then it would be possible to return a null determinant,
rejecting the excitation and attempting again in the next iteration. However, in many cases, this is a common enough
occurrence that it is worth renormalizing the allowed choices of $i$ at the beginning of each single excitation attempt.
Consequently, {\em either} an $\mathcal{O}[N_{\sym}]$ or $\mathcal{O}[N]$ operation can be performed, whichever is smaller,
to calculate the number of electrons in $|D_{\textrm{src}} \rangle$ which have no allowed single excitations, $\delta_s$. This is most
easily calculated from the precalculated symmetry lists, or the electron occupation representation of the determinant respectively,
and is shown in (2) of Fig.~\ref{fig:excit-gen}, but can be omitted in cases with large $\frac{M}{N}$, or no symmetry. 
In rare cases where no single excitations are available from any electron, a null excitation is returned (3).

An electron is then chosen at random (4), with uniform probability of
\begin{equation}
    p(i) = \frac{1}{N}  .
\end{equation}
Generally, choosing electrons is done on the electron occupation representation of the determinant, while testing whether orbitals are
occupied or unoccupied is performed on the bit-string representation. In this way, both tasks can be done in $\mathcal{O}[1]$ time.
Therefore, having both representations available is advantageous. It is then simple to calculate
the number of unoccupied orbitals of the correct symmetry which can be excited to, $M_{\allow}$, via the precomputed symmetry lists. If no
excitations are available from this electron, i.e. $M_{\allow}=0$, another electron can be selected at random if $\delta_s$ has been 
computed previously (5), otherwise a null excitation is returned\footnote{Note that in this case, it is {\em not} possible to simply generate
a double excitation instead, since this possibility would not be reflected in the probability $P_{\textrm{double}}$. Renormalization to account for this
would be possible but inefficient for such an edge case, and so a null excitation should be returned. This issue highlights a common theme in these excitation
generators, where the algorithm attempts to balance the time needed to renormalize the probabilities to allow redrawing if there are no available
excitations of a stochastically chosen type, against the probability of this redrawing needing to occur in the first place. 
From a quality of sampling per iteration perspective, it would be optimal if we could always return an allowed
excitation, but the structure of the algorithm means that it is not always worthwhile to do so.}.

Once orbital $i$ is chosen and $M_{\allow}$ computed, orbital $a$ can be chosen with probability $M_{\allow}^{-1}$ (6). Since it is unknown which orbitals
these $M_{\allow}$ possibilities refer to, and we must ensure that the condition $a \notin |D_{\textrm{src}} \rangle$ is met, we simply draw 
orbitals randomly from all allowed orbitals in the desired symmetry, and test whether they are occupied from the bit-string representation (7). 
It would alternatively be possible to loop through all orbitals until the desired $a$ from the range $M_{\allow}$ allowed orbitals is found, but
this would only be worthwhile for very small $\frac{M}{N}$ ratios, since it would introduce an $\mathcal{O}[M]$ operation.
Finally, once $i$ and $a$ are chosen, the new determinant can be created from the old bit-string representation, the parity change calculated
between the determinants in the standard way, and the orbitals returned to facilitate subsequent calculation of the matrix element between them (8).
The overall probability of generating the new determinant can also be calculated (9), as
\begin{eqnarray}
    P_{\textrm{gen}}(i \rightarrow a) &=& P_{\textrm{single}} \times p(i) \times p(a|i) \times \frac{N}{N-\delta_s}    \\  \label{eqn:p_sing}
    &=& P_{\textrm{single}} \times \frac{1}{N} \times \frac{1}{M_{\allow}} \times \frac{N}{N-\delta_s} \\
    &=& \frac{P_{\textrm{single}}}{M_{\allow}(N-\delta_s)}  ,
\end{eqnarray}
where the last term in Eq.~\ref{eqn:p_sing} is the renormalization factor to account for the number of electrons with no single excitations.

\subsection{Double Excitations}

The logic behind double excitation generation is very similar to that for single excitations, but is a little more involved.
First, a unique $\{i,j\}$ pair is picked uniformly in $\mathcal{O}[1]$ time, with uniform probability
\begin{equation}
p(i,j) = \frac{2}{N(N-1)}   ,
\end{equation}
using an inverse triangular indexing (10). Unlike for single excitations, it was not deemed worth renormalizing in general
for the case of no excitations from a given pair. The point group, angular or linear momentum symmetry of the second unoccupied orbital, $b$,
is uniquely determined by the symmetries of the $\{i,j\}$ pair and $a$, from the requirement to satisfy
\begin{equation}
    \Gamma_a \otimes \Gamma_b = \Gamma_i \otimes \Gamma_j . \label{eqn:sym1}
\end{equation}
When picking the first unoccupied orbital, $a$, the choice of spin must be considered. If both electrons in the $\{i,j\}$ pair have the same spin, the choice
of $a$ must also be constrained to that spin. Otherwise, there are no constraints. This gives the number of possible $a$ orbitals to select,
$M_{\allow}^a$, which is $2M-N$ for a mixed spin $\{i,j\}$ pair, or $M-N_{\sigma}$ for a pair of same spin $\sigma$.
Similarly to the single excitation case, an $\mathcal{O}[N_{\mathrm{sym}}]$ operation is performed to count the number of unoccupied $a$ orbitals, $\delta_d$,
that if picked, would have no symmetry and spin allowed $b$ orbitals with which it could be paired. This allows for 
analytic renormalization of the probabilities, such that multiple attempts at picking $a$ orbitals are allowed if no corresponding
$b$ orbitals are symmetry allowed (11). In the rare case that this number encompasses all unoccupied orbitals of the required 
spin, i.e. $M_{\allow}^a=\delta_d$, a null excitation is returned (12).

From within the spin constraints, a random selection of the first unoccupied orbital can occur. Redrawing of $a$ orbitals from within the $M_{\allow}^a$ set is
allowed if no unoccupied symmetry-allowed $b$ orbital can follow, i.e. $M_{\allow}^{(b|a)} = 0$.
For small or highly symmetric systems, where there are fewer than four possible $a$ orbitals
to successfully pick, the random selection of $a$ is made directly from within the $M_{\allow}^a-\delta_d \leq 3$ range, and 
then an $\mathcal{O}[M]$ operation performed to search
for them (13-14). This ensures that only one random number is required to be drawn to locate an $a$ orbital, rather than potentially drawing many.
This should not be an issue for larger systems with more unoccupied orbitals and less 
symmetry, where an $\mathcal{O}[1]$ operation with infrequent redrawing is used, as the inadvertent selection
of a forbidden or occupied orbital is rarely encountered (15-16). 

Finally, from the conditions in Eq.~\ref{eqn:sym1}, as well as any other symmetry or spin constraints, a random selection of
the $b$ orbital can be made uniformly from the $M_{\allow}^{(b|a)}$ possibilities (17-19). From the final $\{a,b\}$ pair, the number of available $a$ orbitals if
orbital $b$ had been chosen first from the $\{a,b\}$ pair must be calculated, since the pair could have been chosen in either order, 
and in general $M_{\allow}^{(a|b)} \neq M_{\allow}^{(b|a)}$. The changing orbitals, resultant doubly excited determinant and parity change can then be returned
for the calculation of the matrix element (8), while the probabilities follow from,
\begin{widetext}
\begin{eqnarray}
P_{\textrm{gen}}(i,j \rightarrow a,b) &=& P_{\textrm{double}} \times p(i,j) \times [p(a|i,j) p(b|a,i,j) + p(b|i,j) p(a|b,i,j)] \frac{M_{\allow}^a}{M_{\allow}^a - \delta_d} \\
                                      &=& \frac{2 P_{\textrm{double}}}{N(N-1)} \left[ \frac{1}{M_{\allow}^a} \frac{1}{M_{\allow}^{(b|a)}} + \frac{1}{M_{\allow}^a} \frac{1}{M_{\allow}^{(a|b)}} \right] \frac{M_{\allow}^a}{M_{\allow}^a - \delta_d}    \\
                                      &=& \frac{2 P_{\textrm{double}}}{N(N-1)(M_{\allow}^a-\delta_d)} \left( \frac{1}{M_{\allow}^{(b|a)}} + \frac{1}{M_{\allow}^{(a|b)}} \right) .
\end{eqnarray}
\end{widetext}
The overall computational scaling for the calculation of a general excitation is therefore $\mathcal{O}[N_{\mathrm{sym}}]$, after an 
initial $\mathcal{O}[N]$ cost per determinant to set up symmetry lists. 

As an aside, a more flexible excitation generation was also investigated to preferentially sample more important transition probabilities, 
via partial enumeration of excitation subsets, weighting of excitations by 
magnitude of the connecting Hamiltonian matrix elements, a renormalization and subsequent random selection within this subset. Although this 
improves the quality of the sampling with respect to important transitions, the overall acceptance ratios, random errors and scaling of the 
method remained relatively unchanged, and the partial enumeration and generation of the matrix elements was not seen as cost-effective to 
the algorithm. The reason for the relative small improvement in the algorithm is due to the fact that any bias in the generation probabilities
must necessarily be unbiased for in the spawning acceptance criterion (Eq.~\ref{eqn:pspawn}), and so overall spawning rates are not significantly 
affected. Importance sampling may still be important to the algorithm in other guises, but this is likely to involve a transformation in the form 
of the underlying sampled wavefunction\cite{Clark2012}.

\subsection{Spin-coupled function excitation generation}
\label{sec:hphf}

For systems with an even number of electrons, every spin state, $S$, contains an eigenfunction (in the absence of fields or relativistic effects)
with an $M_s$ value of zero onto which the wavefunction can be projected. In this $M_s=0$ sector, there
is an additional time-reversal symmetry which can be imposed on the Hamiltonian matrix, as detailed in Ref.~\onlinecite{BCTA2011}.
Spin-coupled pairs of determinants can be generated by flipping the spin of all electrons, to create pairs of determinants
whose coefficients differ only by a sign-change based on the desired total spin of the system and number of unpaired electrons in the determinant pair,
\begin{equation}
    C_{I_{\alpha}I_{\beta}} = (-1)^SC_{I_{\beta}I_{\alpha}} ,   \label{eqn:EqualCoeffs}
\end{equation}
where $I_{\alpha}$ represents the second quantized string corresponding to the alpha electrons. These pairs of functions are
therefore constructed as 
\begin{equation} 
| \Phi_{IJ} \ket  = \left\{ 
    \begin{array}{l l}
        |I_{\alpha}J_{\beta} \ket & \quad \mbox{if $I = J$}     \\  \label{eqn:SpinCoup} 
        \frac{1}{\sqrt{2}}\left[ |I_{\alpha}J_{\beta} \ket + (-1)^S|J_{\alpha}I_{\beta} \ket \right] & \quad \mbox{if $I > J$}    \\   \end{array}  \right.  
\end{equation}
which constitutes an orthonormal space with the same particle exchange antisymmetry properties as the underlying Slater determinant space, 
but with the symmetry given in \rff{eqn:EqualCoeffs} imposed. 
These time-reversal symmetry functions can be considered an intermediary between determinants and configuration state functions,
and constitute spin eigenfunctions for functions of two unpaired electrons\cite{Helgie,Larsen2000}. 
The performance of FCIQMC in a fully spin-adapted configuration state function space will be considered in a forthcoming paper.

\begin{figure}
\begin{center}
\includegraphics[scale=0.475]{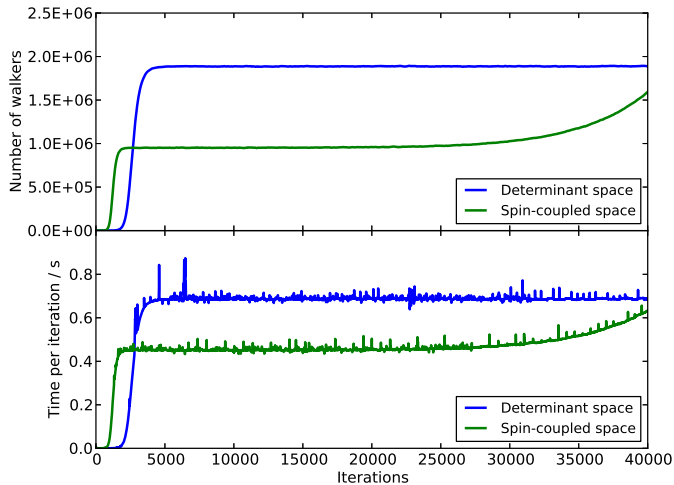}
\end{center}
\caption{The upper panel shows the growth of walkers for a fixed shift FCIQMC calculation (without initiator approximation) for
the equilibrium beryllium dimer in a cc-pVQZ basis\cite{Dunning1989}. The plateau height indicates the number of walkers required 
for annihilation events to control 
the sign problem, and allow convergence to the exact solution. The figure shows that number of walkers required to correctly and 
stably sample this space is half that of the uncontracted determinant space. The cost per iteration is shown in the lower panel, and 
is a factor of $\approx1.5$ cheaper in the spin-coupled space, due to the reduction in walker count somewhat offset by the 
larger spawning rate due to the increased timestep and slightly increased computational overheads. 
}
\label{fig:HPHF1}
\end{figure}

\begin{figure}
\begin{center}
\includegraphics[scale=0.475]{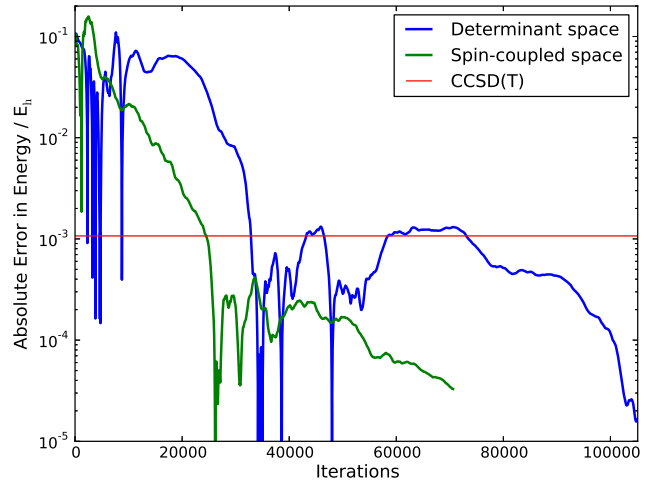}
\end{center}
\caption{Convergence of the error in the energy of the cc-pVQZ beryllium dimer compared to the exact (FCI) energy, in a basis of determinants and spin-coupled functions. 
Due to the reduction in connectivity of the space, the timestep, and hence the rate of convergence with respect to iteration, associated with
the spin-coupled calculation are doubled. The CCSD(T) energy is given for comparison.
}
\label{fig:HPHF2}
\end{figure}

Both the size and internal connectivity of the space of functions shown in \rff{eqn:SpinCoup} are roughly half of those in the 
underlying determinant space. This allows
for fewer walkers to be used to sample the space, and a larger timestep to be used to converge to the solution 
using less imaginary time. This results in an overall approximate 3-4 fold saving 
in computational effort after consideration of the additional overhead for excitation generation and matrix element evaluation, as shown in Figs.~\ref{fig:HPHF1} and \ref{fig:HPHF2}. 
Additionally, the instantaneous walker distribution
is a more spin-pure representation than in the determinant basis, where the condition of \rff{eqn:EqualCoeffs} is only fulfilled in a time-averaged sense.
Separate convergence to the lowest energy states in both odd and even spin sectors of the Hamiltonian is also possible, as the parity of the 
spin-coupled functions will be reversed in each case.

Within the determinantal excitation scheme outlined in section~\ref{sec:Excitgen}, excitations between these spin-coupled functions can be 
easily computed. Although there are up to four possible primitive determinantal couplings between two spin-coupled functions, due to the 
symmetry we only need to consider excitations from one of the constituent determinants in the function, and
{\em define} the excitation probability from the other to be zero. This is possible since all coupled functions are connected from both 
constituent determinants. The fact that the excitation from the chosen primitive determinant may be coupled to both primitive determinants
in the excited spin-coupled function results in the general halving of the the generation probabilities. However, as with the matrix element
evaluation, a small overhead is required for the calculation of this one extra primitive generation probability to correctly modify the 
overall excitation probability.


\section{Application to the Chromium dimer}

We present here an initial study into the chromium dimer, demonstrating the ability of the method to scale to large system sizes and walker numbers,
as well as the ability to be run on many computing cores. This system has been heralded as one of the most challenging in the quantum chemical community, 
with practically all methods struggling with its mix of important dynamical and static correlation effects owing to bonding character between 
both low-lying 3d and more diffuse 4s orbitals, as well as its weak and unusual hextuple bond. Also of importance are the correlation effects of the 3p 
electrons, which have been shown to be crucial to a quantitative description of the binding\cite{Knowles2004}. This suggests that a 
multireference, correlated treatment of 24 electrons is required to correctly capture the physics of the bond. One method which has had 
some success in treating the system is the density matrix renormalization group (DMRG) approach\cite{Yanai2011,Yanai2009,Sharma2012}, and 
it is against this which we benchmark our results here.

\begin{figure}
\begin{center}
\includegraphics[scale=0.475]{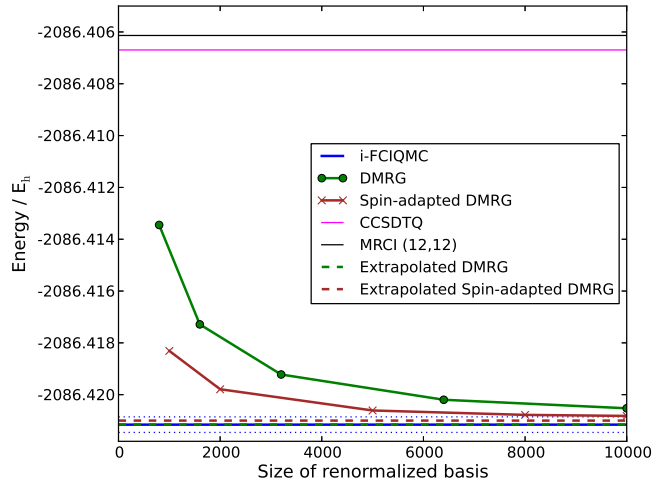}
\end{center}
\caption{Convergence of DMRG with size of renormalized space for the chromium dimer at a geometry of 1.5\AA~in an SV basis, showing agreement in the 
extrapolated limit (dashed lines) with $i$-FCIQMC. 200 million walkers were used. 
DMRG results, as well as those of CCSDTQ and MRCI (from CASSCF(12,12) active space) are obtained from Kurashige et al.\cite{Yanai2009}, while DMRG in 
the spin-adapted basis is obtained from Sharma et al.\cite{Sharma2012} The extent of the $i$-FCIQMC errors bars are given by the dotted lines.
}
\label{fig:Cr2}
\end{figure}

For consistency with the large-scale DMRG study of Kurashige et al.\cite{Yanai2009}, as well as a spin-adapted DMRG study of Sharma et al.\cite{Sharma2012}, 
we performed a single-point calculation of the chromium 
dimer, at a bond length of 1.5\AA, correlating 24 electrons in an orbital space of 30 Hartree--Fock orbitals, formed from an underlying SV 
basis\cite{Ahlrichs1992}. Although this small basis will be insufficient to capture the dynamical correlation effects, it should be a sufficient model 
of the valence space to include the strong static correlation effects of the d-d binding which dominates this bond length. This space is
well outside what could be achieved with FCI, as a single vector of the size of the space would require around 60,000 terabytes 
of memory. By contrast, the $i$-FCIQMC calculation shown in Fig.~\ref{fig:Cr2} was performed with 200 
million walkers ($\approx 3$~gigabytes of distributed memory). After an initial walker 
growth stage performed on small-scale resources, the calculation was run for on 576 cores for 34 hours while statistics were accumulated. 
The resultant energy of \mbox{-2086.4212(3)} Hartrees is in almost exact agreement with the DMRG results extrapolated to infinite bond 
dimension (\mbox{−2086.421156} Hartrees from Ref.~\onlinecite{Yanai2009} and \mbox{−2086.42100} Hartrees from Ref.~\onlinecite{Sharma2012}), 
and gives strong confidence in the value. It is hoped that this accuracy can be maintained while pushing to a correct description of the dynamic 
correlation, and an extension to the rest of the potential surface in the future.

\section{Conclusions}

In this paper, we have detailed the algorithmic aspects of the FCIQMC method as implemented in {\tt MOLPRO} and the freely-available standalone {\tt NECI} 
code\footnote{Standalone code can be cloned from https://github.com/ghb24/NECI\_STABLE.git. Please feel free to contact us for assistance with the compilation and 
running of the program.}, focusing on the parallelism/annihilation scheme and the random 
excitation generation for large-scale calculations. It is hoped that this will be useful to others aiming to implement FCIQMC or 
related methods and those in the lattice DMC and connected communities, as well as beginning a discussion on alternative and optimal algorithms. Other
aspects of the algorithm that were largely omitted, such as error analysis of correlated ratios as required in \rff{eqn:ProjE} or calculation of 
parity changes and matrix elements between determinants are largely covered in other materials\cite{Flyvbjerg1989,Helgie}. In addition, 
the algorithm describes only the bare 
discrete walker dynamics. Moving to a hybrid real-integer representation of the wavefunction has been shown to provide substantial reductions in the size 
of the random errors for equivalent computational effort\cite{Umrigar2012}, and can be easily incorporated into this algorithmic structure. 
The algorithm was also applied to the paradigmatic correlation problem of the chromium 
dimer, where an $i$-FCIQMC calculation was converged to close agreement with large-scale DMRG calculations, and is expected to be at the FCI limit. 
This provides the confidence to apply the method to larger challenging metallic systems where other methods struggle, and combine it with 
complementary approaches for dynamic correlation.

\vspace{5mm}

\section{Acknowledgements}

We would like to sincerely thank Andy May and Peter Knowles for their help and support with the integration of FCIQMC 
into the {\tt MOLPRO} codebase. In addition, we thank Alex Thom and James Spencer for useful discussions, and Jennifer Mohr, Nick 
Blunt, Michael Morkle, Adam Holmes and Cyrus Umrigar for helpful comments on the manuscript. We gratefully acknowledge funding from Trinity College, Cambridge, 
and an EPSRC grant EP/J003867/1.


\end{document}